\documentclass[]{spie}  

\newcommand\thintilde{{\lower.92ex\hbox{\mathtt{\char`\~}}}}

\usepackage{amsmath,amsfonts,amssymb}
\usepackage{graphicx}
\usepackage[colorlinks=true, allcolors=blue]{hyperref}

\title{Optimized green fluorescent protein fused to  F$_\mathrm{o}$F$_1$-ATP synthase for single-molecule FRET using a fast anti-Brownian electrokinetic trap}

\author[a]{Maria Dienerowitz}
\author[a]{Mykhailo Ilchenko}
\author[a]{Bertram Su}
\author[b]{Gabriele Deckers-Hebestreit}
\author[c]{G\"unter Mayer}
\author[c]{Thomas Henkel}
\author[a]{Thomas Heitkamp}
\author[a,d]{Michael B\"orsch}
\affil[a]{Single-Molecule Microscopy Group, Jena University Hospital, Friedrich Schiller University Jena, Nonnenplan 2-4, D-07743 Jena, Germany}
\affil[b]{Department of Microbiology, University of Osnabr\"uck, Barbarastrasse 11, D-49076 Osnabr\"uck, Germany}
\affil[c]{Leibniz-Institut f\"ur Photonische Technologien (IPHT), Jena, Germany}
\affil[d]{Abbe Center of Photonics (ACP), Jena, Germany}

\authorinfo{E-mail: maria.dienerowitz@med.uni-jena.de, Telephone: +49 36419396615, http://www.single-molecule-microscopy.uniklinikum-jena.de}

\begin{document} 
\maketitle

\begin{abstract}
Observation times of freely diffusing single molecules in solution are limited by the photophysics of the attached fluorescence markers and by a small observation volume in the femtolitre range that is required for a sufficient signal-to-background ratio. To extend diffusion-limited observation times through a confocal detection volume, A. E. Cohen and W. E. Moerner have invented and built the ABELtrap - a microfluidic device to actively counteract Brownian motion of single nanoparticles with an electrokinetic trap. Here we present a version of an ABELtrap with a laser focus pattern generated by electro-optical beam deflectors and controlled by a programmable FPGA chip. This ABELtrap holds single fluorescent nanoparticles for more than 100 seconds, increasing the observation time of fluorescent nanoparticles compared to free diffusion by a factor of 10000. To monitor conformational changes of individual membrane proteins in real time, we record sequential distance changes between two specifically attached dyes using F\"orster resonance energy transfer (smFRET). Fusing the \textit{a}-subunit of the F$_\mathrm{o}$F$_1$-ATP synthase with mNeonGreen results in an improved signal-to-background ratio at lower laser excitation powers. This increases our measured trap duration of proteoliposomes beyond 2 s. Additionally, we observe different smFRET levels attributed to varying distances between the FRET donor (mNeonGreen) and acceptor (Alexa568) fluorophore attached at the \textit{a}- and \textit{c}-subunit of the F$_\mathrm{o}$F$_1$-ATP synthase respectively. 
\end{abstract}


\section{INTRODUCTION}
\label{sec:intro}  

Enzymes are protein nanomachines converting or synthesizing small molecules in the cell. We are interested in the fundamental principles of catalytic activity and associated energy consumption as well as conversion processes of these nanomachines. The main focus of our work is the mechanochemistry of the F$_\mathrm{o}$F$_1$-ATP synthase\cite{Borsch1998,Steinbrecher2002,Borsch2002,Borsch2003,Diez2004b,Boldt2004,Steigmiller2004,Diez2004,Krebstakies2005,Zimmermann2005,Steigmiller2005,Zarrabi2005,Zimmermann2006,Zarrabi2007,Zarrabi2007b,Peneva2008,Diepholz2008,Galvez2008,Duser2008,Zarrabi2009,Johnson2009,Duser2009,Modesti2009,Borsch2011,Borsch2011b,Seyfert2011,Ernst2012,Ernst2012b,Hammann2012,Sielaff2013,Borsch2013,Heitkamp2013,Borsch2013b,Bockenhauer2014}. This membrane-bound, large enzyme generates and maintains the chemical energy level of the cells by regulating the concentration of adenosine triphosphate (ATP). F$_\mathrm{o}$F$_1$-ATP synthase is present in all organisms: from bacteria and plant cells to mammalian cells. In the bacterium \textit{Escherichia coli}, the enzymes are driven by a proton concentration difference and an electric potential across the respective membrane. F$_\mathrm{o}$F$_1$-ATP synthase operates as a rotary motor. Therefore, internal parts (the $\gamma$, $\varepsilon$ and \textit{c}-subunits) move stepwise with respect to outer parts (the static $\alpha$, $\beta$, $\delta$, \textit{a}- and \textit{b}-subunits, nomenclature of the bacterial enzyme). 

To measure the speed and step size of rotary subunits and the corresponding catalytic activity, we purify F$_\mathrm{o}$F$_1$-ATP synthase from \textit{Escherichia coli} and re-integrate the enzyme into artificial lipid bilayers. This biochemical technique is called reconstitution. The lipid bilayer is a closed spherical vesicle with a diameter of about 120 nm. Proteoliposomes are lipid vesicles with integrated membrane proteins. Reconstituted F$_\mathrm{o}$F$_1$-ATP synthase is capable of ATP synthesis as well as ATP hydrolysis. 

We monitor subunit rotation by F\"orster resonance energy transfer (FRET). Two different fluorophores are attached to a single enzyme: one at a rotary and a second one at a static subunit. FRET measures the distance change between the two markers, covering a range of 2 to 8 nm during stepwise rotation with millisecond time resolution. Therefore, FRET resolution matches the catalytic turnover of {\raise.17ex\hbox{$\scriptstyle\sim$}}20 to several hundred ATP per second. 

Confocal single-molecule FRET (smFRET) measurements probing freely diffusing proteoliposomes in a buffer droplet are straightforward to implement in a standard microscope setup. The advantage of this method is a high photon count rate with good signal-to-background ratio from individual FRET-labelled enzymes in liposomes. However, the proteoliposomes perform Brownian motion and thus pass the detection volume of the laser focus along random trajectories. The resulting time-dependent intensity trace of FRET-labelled proteoliposomes contains short photon bursts of about 20 to 50 ms and exhibits large intensity fluctuations within a burst. Consequently, the reliability of calculated FRET efficiencies per data point is intensity-dependent and changes within a photon burst.

Therefore, a single molecule manipulation tool is highly desirable for trapping and keeping a FRET-labelled molecule in place in solution. This device has been invented and realised by A. E. Cohen and W. E. Moerner about 10 years ago and is called 'Anti-Brownian Electrokinetic trap' (ABELtrap)\cite{Cohen2005b,Cohen2005c,Cohen2005d,Cohen2007b,Cohen2007,Cohen2008,Jiang2008,Goldsmith2010,Wang2010,Bockenhauer2011,Wang2011,Jiang2011,Wang2012,Bockenhauer2012,Thompson2012,Wang2013,Schlau2013,Bockenhauer2013,Wang2014,Fields2010,Cohen2005,Fields2011}. We have built such a device and show how to hold a single labelled F$_\mathrm{o}$F$_1$-ATP synthase in liposomes in solution using a confocal setup for smFRET. To increase the residence time of the proteoliposome in the ABELtrap, we attached mNeonGreen - a novel, bright green fluorescent protein - to the C-terminus of the \textit{a}-subunit of the enzyme. In addition, this paper presents the first single-molecule FRET data using mNeonGreen as the FRET donor on the F$_\mathrm{o}$F$_1$-ATP synthase.

\section{MATERIALS AND METHODS}
\label{sec:exp}  

\subsection{The ABELtrap Setup}

Our ABELtrap system combines a confocal microscope for single-molecule detection with an FPGA-controlled electrokinetic trap. An extensive description is detailed elsewhere\cite{Fields2011,Su2015}. In essence, a 491 nm cw excitation laser (Cobolt Calypso) passes two orthogonally aligned EODs (Model 310A, Conoptics) that control the beam displacement in the horizontal and vertical direction. A dichroic beam splitter (zt488rdc, AHF T\"ubingen) directs the beam into the 100x microscope objective (Olympus, TIRF, oil, NA 1.49) of an inverted microscope (Olympus IX71) and blocks backscattered light from the sample. 

The detection unit of our system consists of two single-photon counting avalanche photodiodes (SPCM-AQR-H 14, Excelitas). Two synchronised TCSPC cards (SPC150, SPC150N, Becker$\&$Hickl) as well as an FPGA card (PCIe-7852R, National Instruments) record the photon counts\cite{Heitkamp2013,Zarrabi2013,Sielaff2013b}. The latter also records the applied feedback voltages of the electrokinetic trap. After passing a confocal pinhole (\o $150\,\mu m$), the signal spectrally separates at a beam splitter (BS 580i, AHF T\"ubingen) into the FRET donor (500-570 nm) and FRET acceptor (595-665 nm) channel. Interference filters (HQ535/70 and HQ630/70, AHF T\"ubingen) reduce stray light. 

The FPGA card is the active control unit of the ABELtrap regulating feedback voltages, EOD beam displacement and data acquisition in real time. We adapted the original LabVIEW-based program by A. Fields $\&$ A. Cohen\cite{Fields2011} for our experiment. 

\subsection{Sample Preparation}

We fabricate the disposable sample chambers from PDMS (Sylgard 184 elastomer, Dow Corning) off a mask wafer (IPHT, Jena) akin to previous designs\cite{Fields2010,Rendler2011,Su2015}. After thorough cleaning in acetone and plasma etching, the PDMS chip bonds onto a glass coverslip. We dilute 20 nm fluorescent beads (505/515 FluoSpheres, Molecular Probes) in deionised water and fill our PDMS chips with $15\,\mu$l of sample solution. 

We label a genetically introduced cysteine in the \textit{c}-subunit of the F$_\mathrm{o}$F$_1$-\textit{a}-mNeonGreen with AlexaFluor 568 C$_5$-maleimide and reconstitute the double-labelled F$_\mathrm{o}$F$_1$-ATP synthase into preformed liposomes\cite{Heitkamp2016}. Before loading the chip with these fluorescent F$_\mathrm{o}$F$_1$-ATP synthases (proteoliposomes), we add $0.14\%$ polyvinylpyrrolidone (PVP, $\mathrm{M_W}$ 40000, Sigma Aldrich)  to the liposome buffer. This prevents sticking of the proteoliposomes to the sample chamber walls. We mount the microscope coverslip with the PDMS chamber on our inverted microscope and probe the sample with a 100x microscope objective (Olympus, TIRF, oil, NA 1.49). 

\section{RESULTS}

  \begin{figure} [ht]
   \begin{center}
   \begin{tabular}{c} 
   \includegraphics[width=15.7cm]{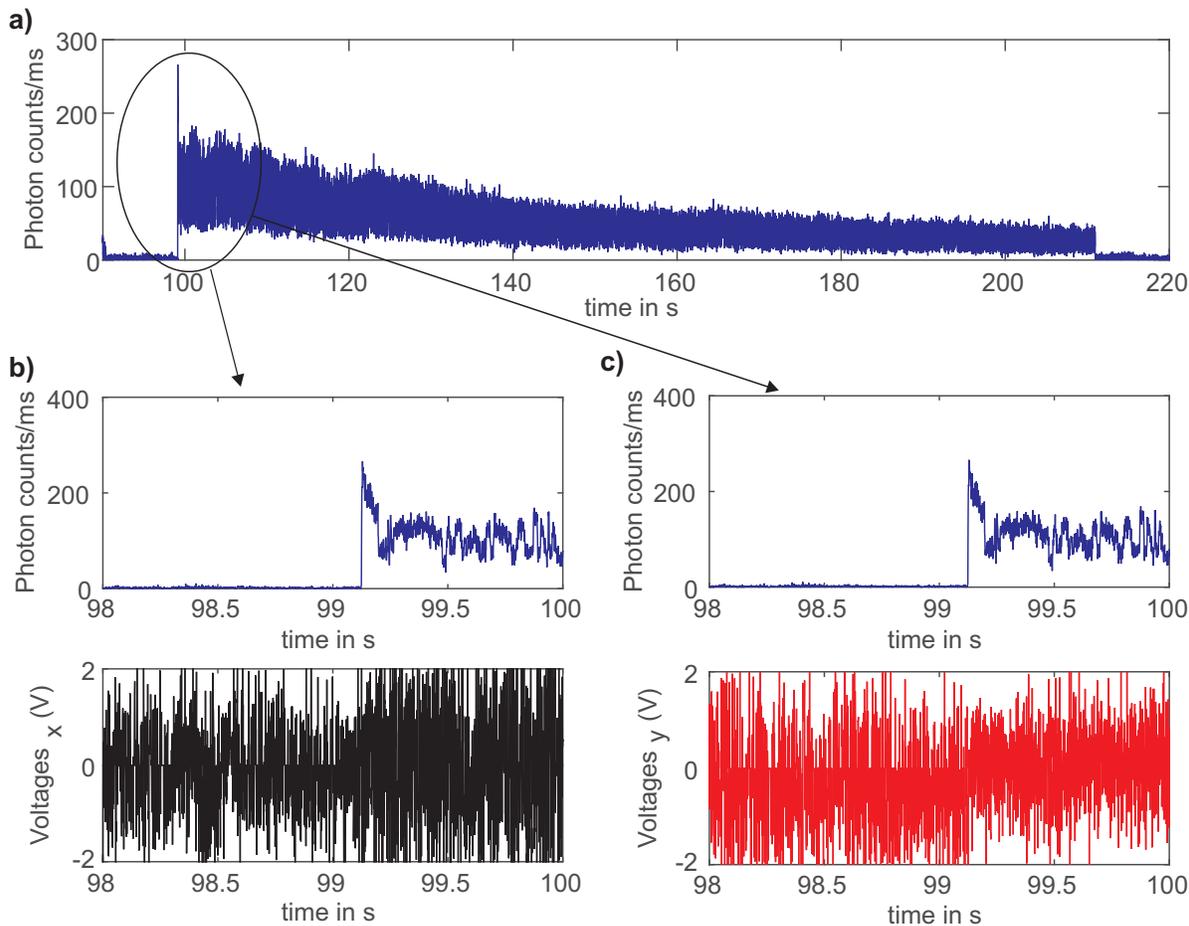}
   \end{tabular}
   \end{center}
   \caption[example] 
   { \label{fig:fig1} 
Electrokinetic trapping of a 20 nm fluorescent bead in an ABELtrap. a) The detected photon intensity increases instantly at the start of a trapping event (@99.1 s) and also drops within one time step once the particle is lost (@210 s). Panels b) and c) display the applied feedback voltages over time in x and y, respectively.}
   \end{figure} 

\subsection{Optimizing electrokinetic trapping for 20 nm fluorescent beads and proteoliposomes}
\label{sec:opti}

  \begin{figure} [ht]
   \begin{center}
   \begin{tabular}{c} 
   \includegraphics[width=15.7cm]{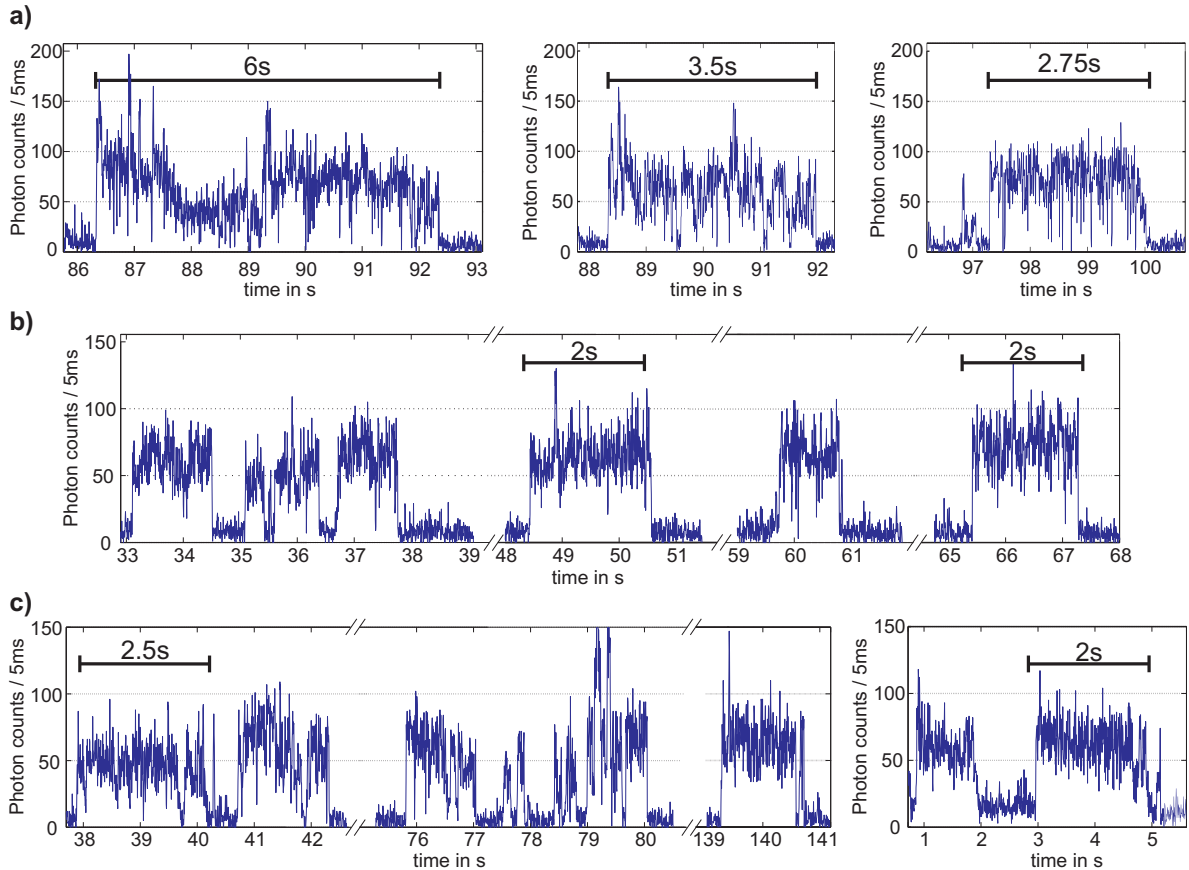}
   \end{tabular}
   \end{center}
   \caption[example] 
   { \label{fig:fig2} 
Electrokinetic trapping of mNeonGreen fused to F$_\mathrm{o}$F$_1$-ATP synthase (FRET donor only sample) in an ABELtrap. Trapping events show a distinct stepwise change in photon count rates at the beginning and end. The constant intensity levels suggest that each proteoliposome contains only one fluorophore. Composed graphs in panel b) and c) originate from the same sample each. Photon count rates are binned in 5 ms time intervals.}
   \end{figure} 

Fluorescent beads are a robust test sample to align and optimize our ABELtrap system. A number of factors significantly alter the performance and hence the ability to trap single nanoparticles or proteoliposomes in the ABELtrap. In our current version of the setup, we optimized the axial position of our sample within the excitation laser focus with a 3D piezo-stage (P-527.3CD with digital controller E-725.3CD, Physik Instrumente, Germany) to improve the signal-to-background ratio. This resulted in trapping events lasting several minutes. 

Figure \ref{fig:fig1}a) shows an intensity trace of a trapped 20 nm fluorescent bead, residing in the trap for 2 minutes. We excited the bead at 491 nm with $3\,\mu W$ laser power. The decay of the photon count rate stems from photobleaching since a single fluorescent sphere hosts a multitude of fluorescent labels. Panels b) and c) show the applied feedback voltages in the x and y direction of the sample chamber. Even if no particle is trapped, the background signal and/or dark counts of the APD cause feedback voltages being applied to the trap site. These counts translate into positions randomly distributed over the entire trapping area, often at opposing points with respect to the trap centre. This results in a frequent sign switch of the applied feedback voltage as displayed in the bottom panel of Figure \ref{fig:fig1}. Once a bead enters the trap site (@99.1 s), the photon intensity instantly increases to more than 100 counts/ms. The feedback voltages stop fluctuating randomly but the mean voltage amplitude still averages to 0 V. Both features indicate a ‘true trapping’ event as opposed to a bead stuck on the sample chamber surface, which would result in non-zero mean feedback voltage. 

  \begin{figure} [h!]
   \begin{center}
   \begin{tabular}{c} 
   \includegraphics[width=15.7cm]{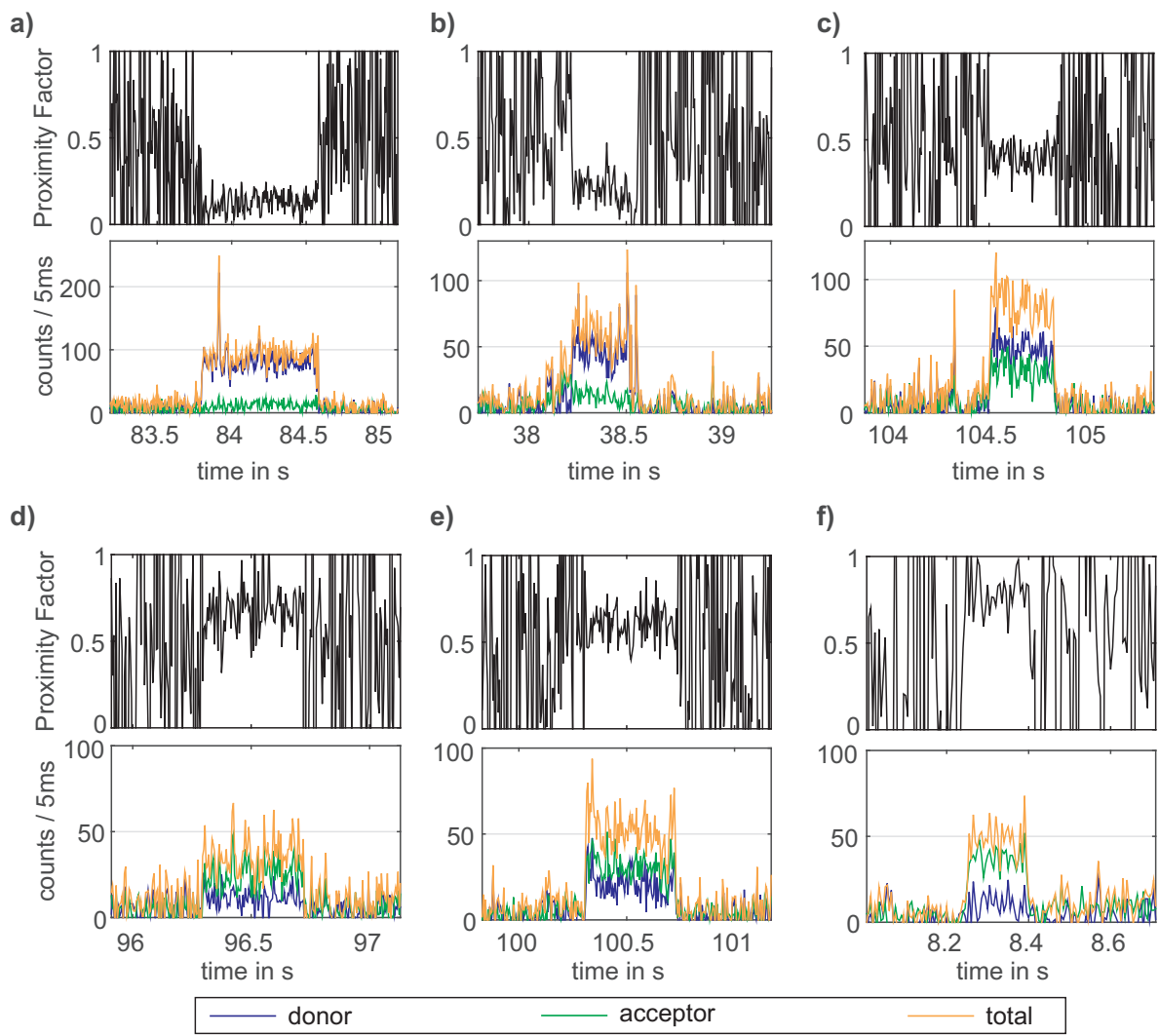}
   \end{tabular}
   \end{center}
   \caption[example] 
   { \label{fig:fig3} 
Intensity traces of FRET incidents in the ABELtrap. a)-f) The top panels show the proximity factor as a measure of different FRET efficiency levels and thus fluorophore distances. The bottom panels display the background subtracted and 5 ms binned intensity traces of the mNeonGreen donor on the \textit{a}-subunit ($I_D$, blue), the Alexa568 FRET acceptor on one \textit{c}-subunit ($I_A$, green) and the total intensity of both (orange). The total intensity is comparable for all trapping events across all samples.}
   \end{figure} 

We previously reported electrokinetic trapping of proteoliposomes containing Alexa488-labelled F$_\mathrm{o}$F$_1$-ATP synthases\cite{Su2015}. To extend the residence time of the proteoliposomes in the trap, we changed the label of the \textit{a}-subunit of the F$_\mathrm{o}$F$_1$-ATP synthase to mNeonGreen\cite{Heitkamp2016}. Figure \ref{fig:fig2} shows sample trapping events of proteoliposomes lasting several seconds. We are thus able to monitor a single proteoliposome over a 100 times longer compared to free diffusion through a laser focus (typically 20-50 ms). On average, our sample contained one or less reconstituted F$_\mathrm{o}$F$_1$-ATP synthase per liposome. This results in less intensity fluctuations per trapped proteoliposome as displayed in Figure \ref{fig:fig2}b). The new mNeonGreen label allowed us to decrease the excitation laser power by a factor of ten (from 130$\mu$W to 30-10$\mu$W). Detecting the trapped proteoliposomes with less laser power reduces background counts and helps to extend the time before the fluorophore photobleaches and thus prolongs the trapping time.

\subsection{Observing single-molecule FRET on the F$_\mathrm{o}$F$_1$-ATP synthase in the ABELtrap}
	
The overall goal for trapping F$_\mathrm{o}$F$_1$-ATP synthases in an ABELtrap is to observe the mechanochemistry of its catalytic activity: ATP synthesis and ATP hydrolysis are associated with subunit rotation and conformational changes within this enzyme. We aim to monitor distance changes between fluorophores attached to rotating subunits as well as non-rotating parts using FRET, while the molecule resides in the trap. We thus have to identify a fluorescent label with an appropriate photostability enabling us to trap the proteoliposomes for extended periods of time. Additionally, we need a matching fluorescent label with a suitable F\"orster radius allowing us to observe single-molecule FRET within a single F$_\mathrm{o}$F$_1$-ATP synthase. 

The trapping experiments of mNeonGreen fused to F$_\mathrm{o}$F$_1$-ATP synthase (see section \ref{sec:opti}) showed promising results in terms of the trapping duration. Next we added an Alexa568 fluorophore to one of the rotating \textit{c}-subunits of mNeonGreen-labelled F$_\mathrm{o}$F$_1$-ATP synthase. The ABELtrap algorithm tracks and acts on the mNeonGreen fluorophore of every proteoliposome. Figure \ref{fig:fig3} shows our first preliminary results of  the double-labelled F$_\mathrm{o}$F$_1$-ATP synthase reconstituted in 120 nm liposomes in our ABELtrap. After subtracting the background counts (6 counts/ms in the FRET donor channel, 4 counts/ms in the FRET acceptor channel), we binned the intensity traces (orange) of the mNeonGreen donor ($I_D$, blue trace), the Alexa568 acceptor ($I_A$, green trace) as well as the total intensity (orange) at 5 ms time intervals. We attribute these low background counts to the optimized excitation laser power ($10\,\mu W$) and the correct positioning of our sample chip along the z direction of the laser focus.

The residence times of the proteoliposomes in the trap are several hundred ms, which exceeds free diffusion times by a factor of 10-20. The proximity factor $P=I_A/(I_D+I_A)$ provides a measure to compare varying FRET levels of different trapping events (see top panels in Figure \ref{fig:fig3} a)-f) ). Low FRET levels correspond to a larger distance between the two fluorescent labels and high FRET levels to a shorter distance. A single \textit{c}-subunit of the F$_\mathrm{o}$F$_1$-ATP synthase is believed to have 10 rotor positions with respect to a static reference point\cite{Jiang2001}, which should theoretically result in the same number of FRET levels. However, due to the symmetry of the 10 \textit{c}-subunits ring, a maximum of only 6 different FRET levels is possible to discriminate\cite{Ernst2012b,Duser2009}. Depending on the signal-to-background ratio, the number of distinct FRET levels could be even smaller. We have so far observed at least 4 different FRET levels associated with the various rotor positions of a single \textit{c}-subunit (see Figure \ref{fig:fig3}).

\section{CONCLUSION}

We presented an EOD-based ABELtrap to hold and monitor single F$_\mathrm{o}$F$_1$-ATP synthases in liposomes. After optimizing the excitation laser power and focus position we were able to trap fluorescent nanoparticles for several minutes. This increases the observation time compared to free diffusion of a 20 nm fluorescent nanoparticle in the confocal detection volume by three to four orders of magnitude. The high signal-to-background ratio allowed us to localise nanoparticles within the given laser focus pattern of our ABELtrap in a fast and precise fashion. Although these low excitation powers lead to minimal photobleaching, we still observe a slow total intensity decay of the trapped nanoparticles. Further improvement of our ABELtrap with respect to detection beam pattern\cite{Wang2010,Wang2013,Wang2014} and tracking algorithm\cite{Wang2011} should increase the residence time as well as the observed intensity fluctuations.

We trapped 120 nm proteoliposomes with single mNeonGreen-labelled F$_\mathrm{o}$F$_1$-ATP synthase for 2 - 6 s. This is up to 100 times longer than the regular free diffusion time. Since a 120 nm proteoliposome diffuses slower than an individual fluorophore, we were able to achieve these long trap durations despite a low signal-to-background ratio of 3 to 5. We aim to further improve this ratio by using all-quartz sample chambers. This will significantly decrease the background signal compared to glass or PDMS chambers as A. E. Cohen and W. E. Moerner have demonstrated\cite{Cohen2008}.

We were able to detect various smFRET levels of trapped double-labelled F$_\mathrm{o}$F$_1$-ATP synthases in liposomes. However, a longer trapping duration would allow us to observe smFRET level changes during catalytic activity, in particular the \textit{c}-subunit rotation. We detected at least 4 different smFRET levels in various photon bursts. The constant overall intensity indicates interactions between two individual fluorophores. Currently, we are exploring smFRET experiments in the presence of ATP to measure ATP hydrolysis in the ABELtrap. Our next goal is applying a proton-motive force to study the subunit rotation during ATP synthesis on trapped single enzymes.

\acknowledgments 
 
We are grateful for the continuous support by Adam E. Cohen and W. E. Moerner. We want to emphasize that Thorsten Rendler, Marc Renz and Anastasiya Golovina-Leiker (Stuttgart, Germany) have built previous EMCCD-based versions of an ABELtrap and provided the design of our PDMS microfluidics. Nawid Zarrabi and Monika D\"user have built the AOD-based confocal ABELtrap (Stuttgart, Germany). We thank Dag von Gegerfelt (von Gegerfelt Photonics, Germany) for the loan of the Cobolt laser, Wolfgang Becker (Becker$\&$Hickl, Germany) for the loan of TCSPC electronics, and Michael Sommerauer (AHF Analysentechnik, Germany) for the loan of high performance optical filters. This work was supported in part by the Baden-W\"urttemberg Stiftung (by contract research project P-LS-Meth/6 in the program "Methods for Life Sciences") and DFG grants BO1891/15-1 and BO1891/16-1 (to M. B.). M.D. gratefully acknowledges funding from the DAAD (Deutscher Akademischer Austauschdienst).

\bibliography{report} 
\bibliographystyle{spiebib} 

\end{document}